\documentclass{article}
\usepackage[utf8]{inputenc}

\usepackage[english]{babel}
\usepackage[utf8]{inputenc}
\usepackage{graphicx}
\usepackage{floatflt}
\usepackage{blindtext}
\usepackage{enumitem}
\usepackage{amsthm}
\usepackage{subfig}
\usepackage{listings}
\usepackage{listingsutf8}
\usepackage{framed}
\usepackage{minibox}
\usepackage{float}
\usepackage{wrapfig}
\usepackage{longtable}
\usepackage[strict]{changepage}
\usepackage{pgfplots}
\usepackage{tikz}
\usepackage{amsmath}
\usepackage{physics}
\usepackage{graphicx}
\usepackage{mathtools}
\usepackage{pdfpages}
\usepackage[version=4]{mhchem}
\usetikzlibrary{matrix}
\pgfplotsset{width=11cm,compat=1.9}
\usepgfplotslibrary{external}
\usepackage[super,comma,sort&compress]{natbib}
\usepackage{hyperref}
\hypersetup{
    colorlinks=false,
     linkcolor=blue,
     filecolor=blue,
     citecolor=blue,
    urlcolor=blue,
}
\DeclareUnicodeCharacter{22C5}{$\dot{}$}
\DeclareUnicodeCharacter{0308}{\"{}}
\usepackage{soul}
\usepackage{authblk}
\usepackage{blindtext}
\makeatletter
\newcommand*{\centerfloat}{%
  \parindent \z@
  \leftskip \z@ \@plus 1fil \@minus \textwidth
  \rightskip\leftskip
  \parfillskip \z@skip}
\makeatother
\usepackage{caption}
\DeclareCaptionLabelSeparator{none}{ }
\addto\captionsenglish{}
\newcommand*{\email}[1]{%
    \normalsize\href{mailto:#1}{#1}\par
    }

\begin{document}
\author[1]{Rosario Roberto Riso}
\author[1]{Tor S. Haugland}
\author[2]{Enrico Ronca}
\author[1,3 *]{Henrik Koch}
\affil[1]{Department of Chemistry, Norwegian University of Science and Technology, 7491 Trondheim, Norway}
\affil[2]{Istituto per i Processi Chimico Fisici del CNR (IPCF-CNR), Via G. Moruzzi, 1, 56124, Pisa, Italy}
\affil[3]{Scuola Normale Superiore, Piazza dei Cavalieri 7, 56126 Pisa, Italy}
\affil[*]{\email{henrik.koch@sns.it}}

\title{\textbf{Molecular orbital theory in cavity QED environments}}

\maketitle

\begin{abstract}
Coupling between molecules and vacuum photon fields inside an optical cavity has proven to be an effective way to engineer molecular properties, in particular reactivity. To ease the rationalization of cavity induced effects we introduce an \textit{ab initio} method leading to the first fully consistent molecular orbital theory for quantum electrodynamics environments. Our framework is non-perturbative and explains modifications of the electronic structure due to the interaction with the photon field. We show that the newly developed orbital theory can be used to predict cavity induced modifications of molecular reactivity and pinpoint classes of systems with significant cavity effects.
We also investigate cavity-induced modifications of molecular reactivity in the vibrational strong coupling regime. 
\end{abstract}

\section{Introduction}
\begin{figure}
    \centerfloat
    \includegraphics[width=1.3\textwidth]{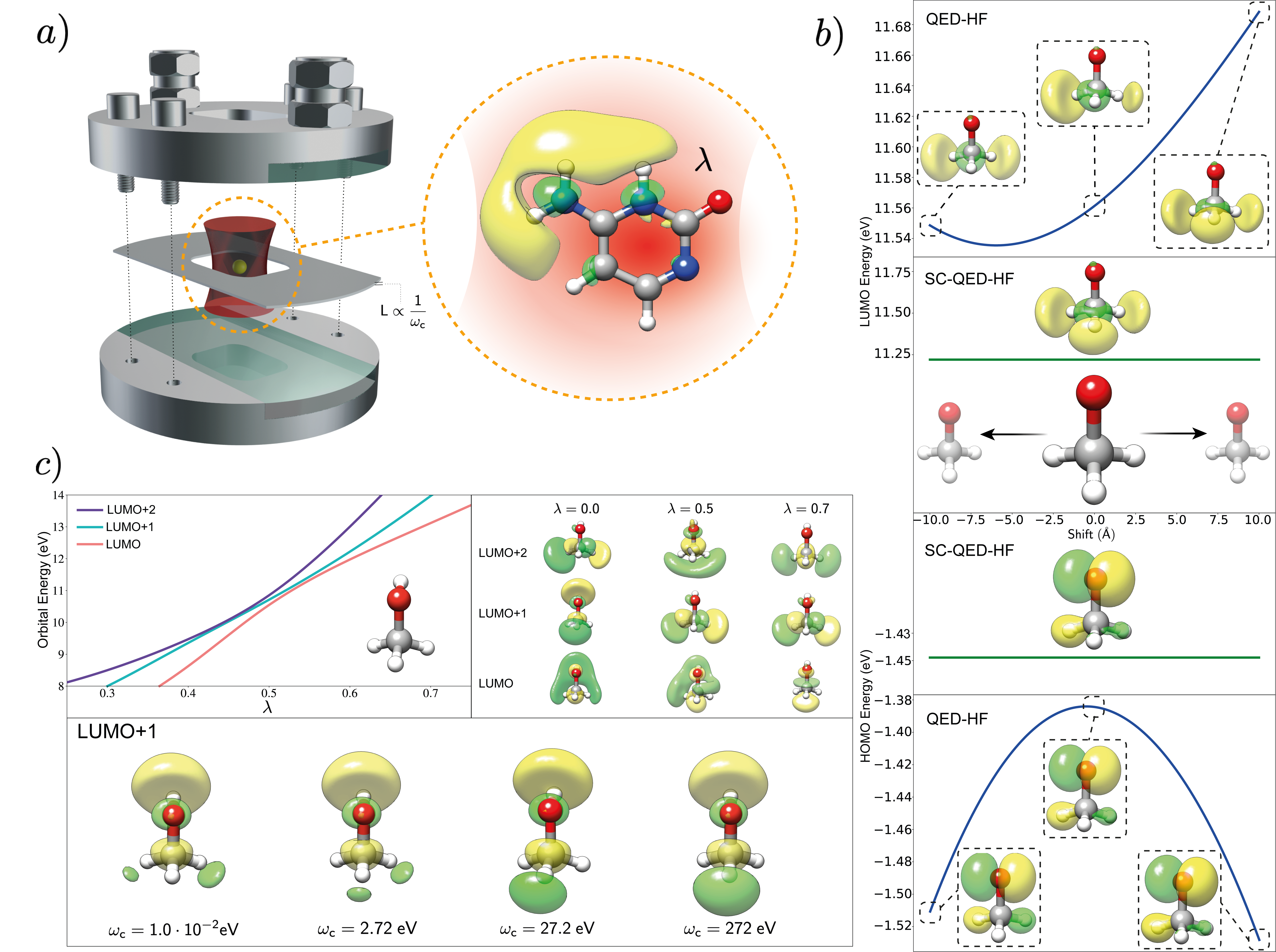}
    \caption{\textbf{Orbitals properties in optical cavities}. a) Pictorial representation of an optical cavity with an injection input, the coated glasses, and a spacer to regulate the distance between the mirrors. The frequency of the cavity fields is proportional to the inverse of the distance between the two mirrors. The lambda parameter quantifies the strength of the light-matter coupling. The cavity picture is based on a video realized by the Ebbesen group. b) Origin invariance of the SC-QED-HF orbitals compared with the origin dependence of QED-HF orbitals. If the methoxy ion is displaced in space, both the orbital energies and the orbital orbital shapes changes for QED-HF orbitals, while the SC-QED-HF orbitals remain unaltered. c) Orbital modifications due to changes in the cavity parameters. Changing the cavity parameters (frequency and coupling) the electronic ground state and therefore the molecular orbitals are changed. Particularly, we show that an avoided crossing like situation can be observed among orbitals.}
    \label{Orbital_properties}
\end{figure}
The use of strong light-matter coupling to modify molecular properties and reactivity is nowadays a very popular topic in physics and chemistry\cite{Polaritons_Scholes,Quantum_Wilson,Cavity_Hagenmuller,Cavity_Wang}.
Many groundbreaking works have shown that interaction with confined fields can impact many matter features ranging from the modification of absorption and emission spectra \cite{Absorption_Herrera,Coherent_Wang,Cavity_Li.T, Polariton_DelPo}  to the alteration of photochemical processes \cite{Modyfying_Hutchison,Investigating_Mandal}.
In particular, the chemistry community has focused on how and to which extent reactions can be engineered by coupling with light \cite{Novel_Bennett,Cavity_Galego, Coherent_Shalabney,Groundstate_Thomas}. 
Interaction with quantum fields can indeed significantly affect molecular processes both in the ground state and excited states\cite{Modifying_Sau,Many_Galego} with reported examples of reactions slowing down \cite{Groundstate_Thomas}, speeding-up \cite{Cavity_Lather} or becoming selective towards one product \cite{Tilting_Thomas}.
The easiest way to achieve strong coupling between light and matter is through optical cavities (See Fig.\ref{Orbital_properties}a), where the frequency of the electromagnetic field 
is determined by the geometrical features of the device \cite{Laser_Siegman}. 
Inside the cavity, the photonic vacuum couples to the molecular system creating polaritonic states  \cite{Light_Wang} with distinct features \cite{Cavity_Herrera}.
Most importantly, the properties of the mixed matter-photon states can be engineered tuning the photonic part of the system, which means that polaritons represent a very effective way to modulate matter properties in a non-invasive way \cite{Polariton_Yuen}.\\
A detailed theoretical description of the strong coupling regime is urgently needed to develop an intuitive picture of cavity chemistry, that would also greatly ease the experimental design.
However, this is a challenging
task because in the strong coupling regime photons become a critical component of the quantum system and they must therefore be treated as quantum particles following quantum electrodynamics (QED).
Only recently, a wider interest from the chemistry 
community in electron-photon systems has lead to the 
introduction of semi empirical methods \cite{Manipulating_Fregoni}, 
variational theories \cite{Variational_Rivera} as well as many others \cite{Polarized_Mandal}. 
Only four \textit{ab initio} methods have been proposed so far: QED Hartree Fock (QED-HF) \cite{Coupled_Haugland}, quantum electrodynamical density functional theory (QEDFT)\cite{Atoms_Flick,Quantum_Ruggenthaler,Ab_Flick},QED coupled cluster (QED-CC) \cite{Coupled_Haugland} and QED full configuration interactions (QED-FCI) \cite{Intermolecular_Haugland,Coupled_Haugland}. 
Molecular orbital theory for the strong coupling regime has so far not been proposed, although all the methods mentioned above use an orbital basis to parametrize the wave function \cite{Coupled_Haugland,buchholz2020light,nielsen2018dressed}. 
The molecular orbital (MO) concept is a powerful theoretical tool
used to develop correlated theories and to provide a qualitative and simple interpretation of molecular properties \cite{Molecular_Fujimoto}.
In particular, since MOs also display local properties of the system like the electrophilicity or nucleophylicity of atomic fragments, reactivity can in many cases be more easily rationalized in terms of orbitals, rather than with electron densities. The introduction of a MO theory for systems in QED environments may therefore significantly enhance our chemical intuition about experiments in strong coupling regimes.\\
In standard quantum chemistry, molecular orbitals are obtained as solutions to the HF equations, however a proper extension of this concept to QED environments has proven to be non-trivial. Indeed, the orbitals provided by QED-HF display unphysical behaviour for charged systems with both the orbital energies and the shapes changing when the molecule is translated (See Fig.\ref{Orbital_properties}b).
The orbital construction scheme is therefore not reliable with critical quantities like the HOMO-LUMO gap \cite{Tuning_Teunissen} being ill-defined.
Moreover, standard perturbation theory to describe electron-electron and electron-photon correlation can not be used without a proper orbital theory\cite{Coupled_Haugland}.  
In this work, we introduce a new \textit{ab initio} electron-photon framework called strong coupling quantum electrodynamics Hartree Fock (SC-QED-HF) which naturally leads to the first fully consistent MO theory in QED environments.
Our analysis reveals that in order to obtain a well behaved molecular orbital description the photonic field contribution must be accounted for in a non-perturbative way, ensuring that in the limit of infinite coupling the exact wave function is obtained.
The newly developed method captures electron-photon correlation allowing us to display field induced effects on the electronic ground state while keeping the computational cost lower than post-HF approaches.
\section{Results}
The photon-matter interaction is described through the well known dipole Hamiltonian \cite{Resolution_Di,Cavity_Ashida,Resolution_Taylor}
\begin{equation}
H = H_{e}+\omega b^{\dagger}b -\lambda \sqrt{\frac{\omega}{2}}\left(\mathbf{d}\cdot\boldsymbol{\epsilon}\right)\left(b+b^{\dagger}\right)+\frac{\lambda^{2}}{2} \left(\mathbf{d}\cdot\boldsymbol{\epsilon}\right)^{2},  
\label{Hamiltonian}
\end{equation}
where $H_{e}$ is the electronic Born-Oppenheimer Hamiltonian, $\mathbf{d}$ is the molecular dipole operator,   $\lambda=\sqrt{\frac{1}{\epsilon_{0}V}}$ is the coupling constant between the molecule and the field, where $V$ is the quantization volume. The $\omega$ and $\boldsymbol{\epsilon}$ are the frequency and polarization of the electric field, and the bosonic operators $b$ and $b^{\dagger}$ respectively annihilate and create photons. 
Besides $H_{e}$, the dipole Hamiltonian is composed of three main terms: the purely photonic term, an interaction term and the dipole self energy $\frac{\lambda^{2}}{2} \left(\mathbf{d}\cdot\boldsymbol{\epsilon}\right)^{2}$. The self energy is often
neglected, however this term is critical to ensure the Hamiltonian is bounded from below and displays the correct scaling with the size of the system \cite{Light_Rokaj}. 
To describe a molecule in a cavity, we must solve the eigenvalue problem for the dipole Hamiltonian
\begin{equation}
H\ket{\psi}=E\ket{\psi},    
\end{equation}
where $\ket{\psi}$ is the electron-photon wave function. 
Within the SC-QED-HF approach the wave function is approximated as
\begin{equation}
\ket{\psi}= \textrm{exp}\left(-\frac{\lambda}{\sqrt{2\omega}}\sum_{p\sigma}\omega_{p}a^{\dagger}_{p\sigma}a_{p\sigma}\left(b-b^{\dagger}\right)\right)  \ket{\text{HF}}\otimes\ket{0_{\text{ph}}}. \label{Guess} 
\end{equation}
Here $a^{\dagger}_{p\sigma}$  and $a_{p\sigma}$ respectively create and annihilate an electron in the orbital $p$ with spin $\sigma$, $\ket{\text{HF}}$ denotes an electronic Slater determinant and $\ket{0_{\text{ph}}}$ is the photonic vacuum. The electron-photon correlation basis is obtained diagonalizing the operator ($\mathbf{d}\cdot\boldsymbol{\epsilon}$). Similar bases were also employed by Huo et al. \cite{Polarized_Mandal} as well as by Ashida et al. \cite{Cavity_Ashida} and Sch{\"a}fer et al. \cite{Schafere2110464118}. However, in our approach the $\omega_{p}$ parameters are variationally optimized in this way dramatically improving the description of the system (See Supplementary Information).
Using SC-QED-HF we obtain origin invariant and size-extensive energies and, most importantly, we can construct fully origin invariant molecular orbitals, see Fig \ref{Orbital_properties}b. Non size-extensive effects in the cavity can be modeled and by optimizing the correlation basis and including more photons in the reference wave function in Eq.\ref{Guess}.  This modification could be crucial if long range collective effects need to be modeled. 

As displayed in Fig.\ref{Orbital_properties}c, both frequency and coupling variations can induce significant modifications of the MOs.
Both $\lambda$ and $\omega$ can be varied by changing the geometrical features of the cavity -- opening the possibility of engineering ground state molecular properties by tuning the field parameters. 
We point out that while the coupling is intrinsically a light-matter property, the frequency is a field property only. The frequency dependent orbitals of SC-QED-HF are dressed with the photon field according to
\begin{equation}
\tilde{a}^{\dagger}_{p\sigma} = \sum_{q} a^{\dagger}_{q\sigma}\;\textrm{exp}\left(\frac{\lambda}{\sqrt{2\omega}}\;\omega_{q}(b-b^{\dagger})\right)U_{qp}, \label{Dressed}
\end{equation}
where $\mathbf{U}$ is the unitary transformation from the HF basis to the correlated basis. 
Both occupied and unoccupied orbitals are affected by the cavity parameters, however, larger changes are typically observed for the latter.
Molecular orbitals are used to qualitatively predict and interpret molecular excitations as well as bond formations, therefore the cavity-induced modification of the orbital character reported in Fig.\ref{Orbital_properties}c suggests significant changes in the molecular properties. 
Orbital avoided crossings and orbital mixing are other effects that can be induced by coupling to the photon field. In particular, when two orbitals have the same symmetry, an “avoided crossing” like situation can be observed (the orbitals mix but their energies never become degenerate, see Fig.\ref{Orbital_properties}c. Conversely, when two almost degenerate orbitals have different symmetry, a crossing is allowed and the orbital ordering is simply exchanged, see Supplementary Information. 
Whenever the HOMO-LUMO gap is small we expect to see particularly large cavity effects and to achieve significant control over molecular properties. In these cases the near degeneracy between the HOMO the LUMO can be enhanced or lifted by the cavity fields \cite{Coupled_Haugland} with significant impact on the ground state wave function. A reliable scheme to model these kinds of systems might require combining the SC-QED-HF Hamiltonian with an active space method \cite{helgaker2014molecular}. 
 \\

\begin{figure}
    \centerfloat
    \includegraphics[width=1.3\textwidth]{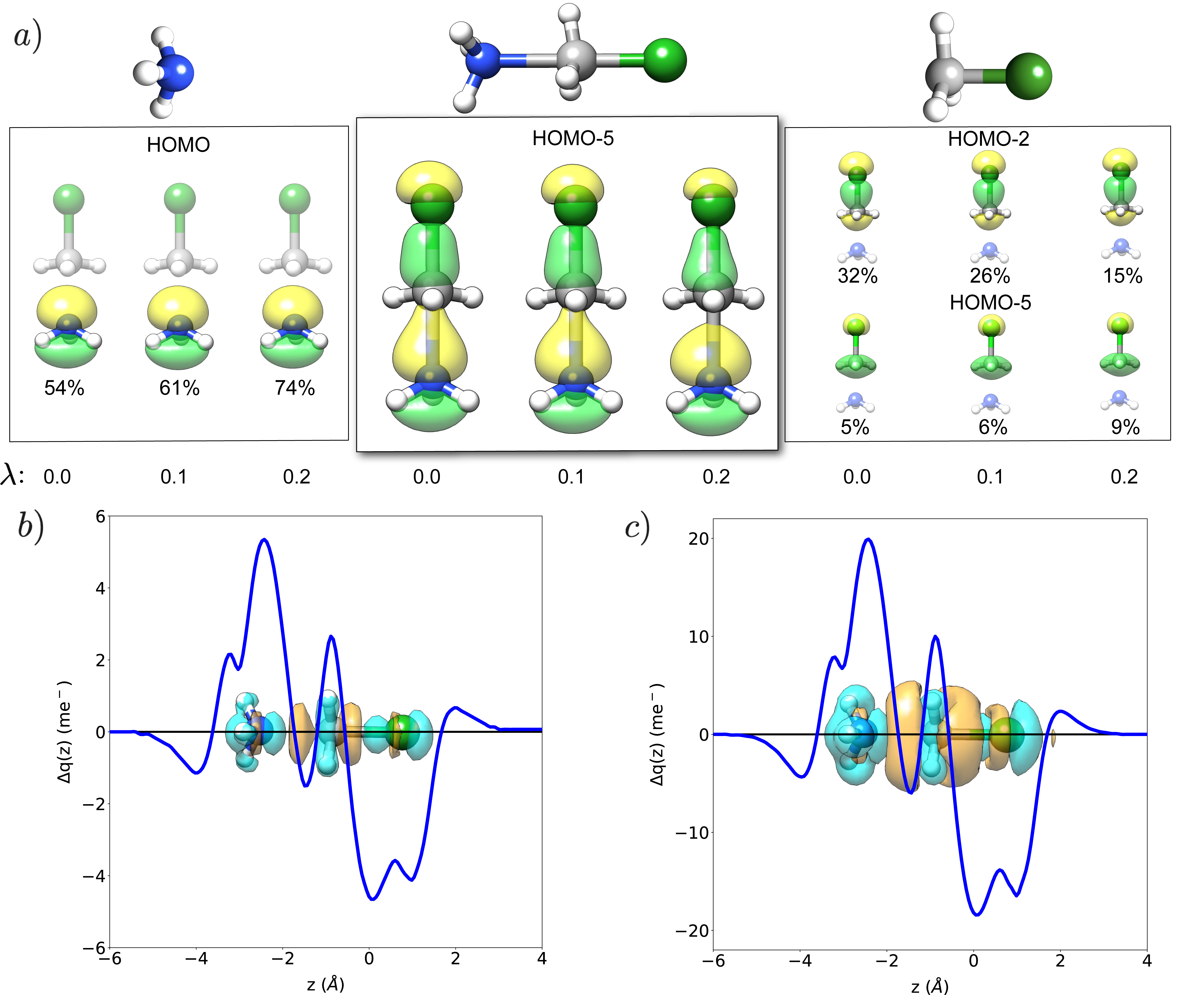}
    \caption{\textbf{Orbital rationalization of molecular reactivity}. a) Orbital analysis of the reaction between ammonia and methyl chloride. Our discussion focuses on the analysis of the only occupied orbital having bonding character on both reagents, the HOMO-5 orbital. We notice that as the coupling increases the bond between carbon and nitrogen becomes weaker. In particular, for coupling $\lambda=0.2$ the HOMO-5 is a simple composition of the non-bonding ammonia lone pair and the \ce{C-Cl} sigma bond. This is verified using a projection of HOMO-5 on the orbitals of the fragments. The contributions are given in percentages.
    b)/c) Difference between electron densities computed at coupling $\lambda=0.1$ /$0.2$ and $\lambda=0.0$. 
    The blue regions of the isosurfaces show charge accumulation inside the cavity while the orange regions show charge depletion. In dark blue we plot the integral of the density differences to further display cavity induced charge reorganization \cite{Coupled_Haugland,Intermolecular_Haugland}. 
    The polarization of the field is placed along the \ce{C-Cl} bond and the cavity frequency is 13.6 eV. }
    \label{Prototype_reaction}
\end{figure}
One of the main applications of molecular orbital theory is the rationalization of molecular reactivity. Specifically, MOs are used to predict the critical steps of a reaction, as well as the effects that a modification of the  conditions can induce on the final outcome. 
Here we investigate whether the molecular orbitals provided by SC-QED-HF can be used to predict cavity effects on the formation of a complex from its initial constituents.
As a case study, we chose a prototypical example of a nucleophilic substitution (SN2): a methyl chloride reacting with an ammonia molecule to produce methylamine and hydrogen chloride.
A particularly interesting configuration along the reactive path is the one reported in Fig.\ref{Prototype_reaction} where both the \ce{C-Cl} and the \ce{C-N} bonds are not fully formed nor broken. Since this arrangement is similar to the transition state of the SN2 reaction, its cavity induced stabilizations or destabilizations 
play a critical role in determining a reaction speedup or slowdown.
Analyzing the MOs of the complex we notice that there is only one occupied orbital with a significant bonding component over both the chloride and the nitrogen, HOMO-5. 
The HOMO-5 orbital of the complex is composed of the nitrogen lone pair, the \ce{C-Cl} sigma bond and the three \ce{C-H} sigma bonds. Projection on the orbitals of the fragments allows us to monitor field induced changes in the character of HOMO-5. 
Specifically, in Fig.\ref{Prototype_reaction} we display how HOMO-5 and its projections change for three different values of the $\lambda$, 0.0, 0.1 and 0.2. Couplings values of 0.1 or 0.2 are extremely high, far superior to what is nowadays experimentally achieved for molecules interacting with a single mode of the quantum field; however, larger values are used here to approximate multimode and collective effects \cite{Intermolecular_Haugland} (See methods).
As the coupling increases, HOMO-5 becomes more localized on the two different fragments. Specifically, the sigma bond between the nitrogen and the carbon disappears as observed from the orbitals of the fragments. Indeed their overlap slightly decreases with the coupling. We notice in particular that for $\lambda = 0.2$, the HOMO-5 orbital has the character of a sigma bond between the carbon and the chloride plus a non-bonding lone pair from the nitrogen. The orbital analysis therefore suggests that cavity induced effects make the configuration in Fig.\ref{Prototype_reaction} less favourable inside the cavity than outside. This statement is confirmed using an energy argument with the formation energy of the complex ($\Delta E(\lambda) = E_{complex}(\lambda)-E_{fragments}(\lambda)$) increasing by $5.30 $ kJ/mol for $\lambda =0.1$ and $17.36$ kJ/mol for $\lambda = 0.2$ compared to outside the cavity.\\
In panels b and c of Fig.\ref{Prototype_reaction}, we display the difference between the ground-state electron densities computed with and without the cavity for different couplings ( $\Delta\rho = \rho_{\lambda}-\rho_{nocav}$, $\lambda = 0.1, 0.2$). In particular, the isosurface shows changes of the electronic density due to the cavity. 
To further analyzed the charge readjustment we report an integration of the density difference along the field polarization in Fig.\ref{Prototype_reaction} \cite{Intermolecular_Haugland, Coupled_Haugland}. 
We observe that as the coupling increases there is a significant charge depletion among both the \ce{C-Cl} and the \ce{C-N} bonds, as expected from the orbital analysis.
This behaviour confirms the conjecture presented in Refs.[\kern-.4em\citenum{Coupled_Haugland,Light_Rokaj,Ab_Flick}] that 
upon interaction with the field, electrons reorganize to minimize the electronic dipole component along $\boldsymbol{\epsilon}$.\\

\begin{figure}
     \centerfloat
    \includegraphics[width=1.3\textwidth]{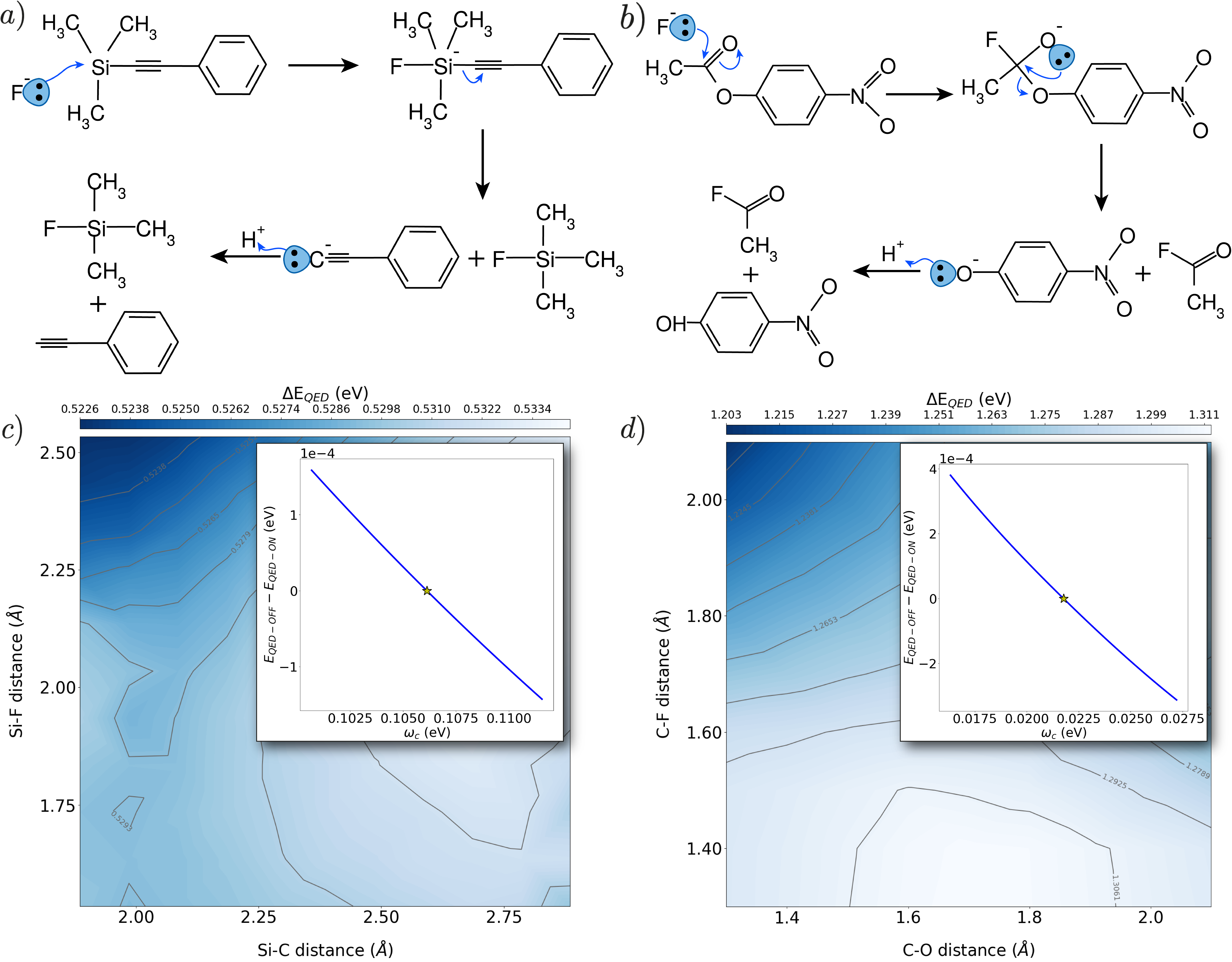}
    \caption{\textbf{Potential energy surfaces for reactions in VSC regime}. Schematic representations of the reaction mechanisms for a) the deprotection of PTA using TBAF and b) the solvolysis of PNPA. The electron movement is represented by blu arrows. c)/ d) Difference between the potential energy surfaces for the PTA deprotection/PNPA solvolysis computed with SC-QED-HF and HF. The dark regions are stabilized by the cavity induced effects while bright regions are destabilized. On the top corner we show that the energy for the PES configurations is monotone in the cavity frequency and therefore no resonance effects are observed. The coupling parameters are $\lambda =0.025 $  and $\lambda = 0.044$ respectively, while the frequencies are in resonance with the \ce{Si-C} and the \ce{C=O} vibrational frequencies. In the insets of panel c) and d) the resonance features of SC-QED-HF are investigated. The yellow star indicates the resonant frequency. The QED effects have been shifted so that at resonance they are equal to zero.}
    \label{fig:Ebbesen_reactions}
\end{figure}
This charge reorganization might actually play a
role in determining cavity induced modifications of molecular reactivity. This idea was also reported by Galego et al. \cite{Many_Galego} and tested using a transition state method for the strong light-matter coupling case.
Strong coupling between molecular vibrations and cavity fields, vibrational strong coupling (VSC), has proven to be a particularly effective way to affect the outcome of molecular reactions  \cite{Groundstate_Thomas,Cavity_Lather,Tilting_Thomas,Cavity_George}. Although the number of experiments displaying to what extent reactions can be engineered through VSC is constantly increasing, several
aspects still remain unclear. 
Indeed, while the formation of vibrational polaritons 
certainly contributes to determine the observed effects, questions still remain on the role played by the electrons and the solvent \cite{Groundstate_Thomas,On_Climent}. Moreover, it has been suggested that an important part might be played by collective effects \cite{Polaritonic_Sidler} and by dark states \cite{Can_Du}.
Just recently, multiple works have suggested that the reaction slowdown in the VSC regimes can be described through nuclear effects only. Explanations use either the dynamical caging of the photon fields \cite{li2021theory} or cavity induced energy redistribution among normal modes \cite{Shining_Schafer}. However, in both cases electron-photon interaction was approximated preventing an accurate quantification of the cavity induced electronic effects on the chemical reactivity. 
Since SC-QED-HF captures some electron-photon correlation we will use it to assess the relevance of the electronic changes in VSC.
We computed the potential energy surfaces (PES) at the HF and SC-QED-HF level of theory for two different reactions that are known for respectively being slowed down and 
catalyzed under vibrational strong coupling conditions: the deprotection of 1-phenyl-2-trimethylsilylacetylene (PTA) with tetra-n-butylammonium fluoride (TBAF)\cite{Groundstate_Thomas} ( Fig.\ref{fig:Ebbesen_reactions}a) and the solvolysis of para‐nitrophenyl acetate (PNPA) with TBAF\cite{Cavity_Lather} (Fig.\ref{fig:Ebbesen_reactions}b). The field polarization was aligned along the \ce{Si-C} and the \ce{C-O} bonds respectively.
A first 
inspection of the potential energy surfaces (See  Supplementary Information) reveals that for both reactions the overall shape is not changed by the cavity. Specifically, in the case of PTA-deprotection an intermediate is formed before the Si-C bond is broken (second step in Fig.\ref{fig:Ebbesen_reactions}a) and a transition state barrier must be overcome. Conversely, in the case of the PNPA-solvolysis the reaction is concerted.
In Fig. \ref{fig:Ebbesen_reactions} we show the difference between the SC-QED-HF and HF potential energy surfaces.
The colour maps in Fig.\ref{fig:Ebbesen_reactions}c) and Fig.\ref{fig:Ebbesen_reactions}d) visualize the QED effects highlighting which configurations are stabilized by the cavity and therefore whether we should expect a reaction slowdown or a speedup.  In particular, the dark zones are stabilized and the bright zones being destabilized upon coupling to the photons.
In both the analyzed cases the reagents are more stabilized than the products pointing towards an overall slowdown of both the reactions.  
In the PTA-deprotection case our prediction is in agreement with what is experimentally observed, while in the case of PNPA, electronic effects would predict a slowdown instead of a speedup. 
Most importantly, however, for the two reactions in Fig.\ref{fig:Ebbesen_reactions} the QED modifications of the electronic structure are so small they do not seem to affect in a significant way the shape of the PES. Stabilizations reach a maximum of 0.01 eV in the case of the PTA-deprotection and to about 0.07 eV in the PNAN-solvolysis case. Moreover, we point out that rate changes in the experiments are only observed if resonance between the cavity mode and the nuclear vibrations is achieved, while similar cavity induced modifications of the PESs are observed in a wide range of cavity frequencies (top-right corner of Fig.\ref{fig:Ebbesen_reactions}c,d). In conclusion the observed cavity induced electronic effects are not large enough to justify the experimental outcome. In particular we highlight that no resonance condition is found in the electronic structure properties. We therefore conjecture that electronic effects alone are of secondary importance in the vibrational strong coupling regime and that no significant variations of the reaction mechanism or of the transition state should be expected for the reported examples in disagreement with what was reported by Galego et al. \cite{Cavity_Galego}. Summarizing, our studies show that electronic effects can be relevant under the appropriate experimental conditions (see Fig.\ref{Prototype_reaction}, near degeneracy systems) but do not seem to play an important role in resonant VSC experiments. 
This is reasonable considering that nuclear properties are only indirectly related to electronic effects.

\section{Conclusion}
In this work we have introduced a non-perturbative electronic structure method for the strong coupling regime, strong coupling QED-HF. The approach effectively captures part of the electron-photon correlation and scales as $N^{3}$ using an optimal implementation. Moreover, it leads to the first fully consistent molecular orbital theory for QED environments. The method is completely general and can therefore be applied to any kind of quantized field environment \cite{fregoni2021strong}. The obtained orbitals have the correct fermionic/bosonic statistics and are dressed with the photonic field as shown in Eq. \ref{Dressed}. Furthermore, in the infinite coupling limit the SC-QED-HF wave function is an exact eigenfunction of the light-matter Hamiltonian. Similarly to standard molecular orbitals, the dressed orbitals can be used to rationalize molecular reactivity. 
From the considered examples we confirm that electrons interacting with the quantized field get more localized and, in this way reduce the electronic dipole in the electric field direction.
The importance of cavity induced electronic effects in vibrational strong coupling can also be investigated using the new method. Our study shows that QED effects on the electronic ground state can be significant under the proper experimental settings (see Fig.\ref{Prototype_reaction}), however they only play a secondary role for the analyzed  VSC cases. Therefore a cavity induced change of the transition states in Fig.\ref{fig:Ebbesen_reactions} is unlikely; in accordance with the conclusions reported in other works \cite{On_Li, li2021theory}.
Furthermore, SC-QED-HF opens the way for developing perturbative methods to capture additional electron-electron and electron-photon correlation with a lower computational cost. This should in particular allow us to describe non size-extensive effects in cavities that are critical to reproduce strong coupling modifications of intermolecular interactions \cite{Intermolecular_Haugland}. Finally, the introduction of a molecular orbital theory opens to the possibility of developing multi-level methods to simulate strong light matter coupling for large molecular systems and include collective effects in an \textit{ab initio} framework.
\section{Methods}
The SC-QED-HF approach is a fully \textit{ab initio} electron-photon correlated framework.
The method has been implemented in a development version of the eT program \cite{An_Folkestad}. 
The optimization of the wave function is performed using the gradients of the energy with respect to the $\omega_{p}$ parameters and the density matrix. The gradients are computed analytically and their explicit expression is reported in the Supplementary Information. The DIIS extrapolation procedure is adopted to speed up convergence \cite{helgaker2014molecular}. The electronic density is initialized as a superposition
of atomic densities while the starting guess for the $\omega_{p}$ parameters is chosen to equal the eigenvalues of $\mathbf{d}\cdot\boldsymbol{\epsilon}$. We have tested other starting values of the $\omega_{p}$ parameters, however the final result is not affected by the initial guess. 
SC-QED-HF allows us to incorporate some electron photon correlation in the wave function without invoking the complexity of methods like coupled cluster or full configuration interaction. In particular, the current implementation of SC-QED-HF scales as $N^{5}$ due to the calculation of the two electron integrals in the dipole basis. Future optimization of the method, including screening of the two electron integral, will reduce the scaling to $N^{3}$.
Additional theoretical details regarding SC-QED-HF are reported in the Supplementary Information. 
All geometries shown in this paper have been optimized using the ORCA package \cite{The_Neese}.
We used DFT-B3LYP/def2-SVP basis set to optimize the geometries for the small molecules (methoxyde, methanol) as well as for the structure in Fig.\ref{Prototype_reaction}. For the optimization of the complex in Fig.\ref{Prototype_reaction} the distance between the nitrogen and the carbon was fixed to 2.1 \AA$\;$while the positions of all the other atoms were reoptimized. The fragments in Fig.\ref{Prototype_reaction} were obtained from the geometry of the complex without any additional optimization. This way we isolated cavity induced effects neglecting modifications of the orbitals induced by the geometry relaxation. Finally, the complex formation energy $\Delta E$ was computed from the fragments energies at infinite separation.
The geometries for the PNPA potential energy surfaces were optimized using DFT-PBE0/def2-SVP while the PTA geometries were computed using  DFT-B3LYP/6-31G* similarly to what is already reported in literature \cite{Shining_Schafer}.
All the calculations reported in this paper have been performed using a cc-pVDZ basis set.
As already reported in the main text, some of the calculations have been performed for large values of the light-matter coupling. This choice is justified because in real cavities molecules interact with more than one mode at the time and that the net effect can be approximated by a larger value of the single mode coupling to the molecule. In particular, in the QED-HF case it can be shown that a single mode calculation is identical to a multimode one with an effective coupling $\lambda^{2}_{eff}=\sum_{i}\lambda^{2}_{i}$. This relation does not hold exactly for SC-QED-HF because of electron-photon correlation, however the net effect still remains an increase of the light matter coupling. \\
\subsection*{Acknowledgements}
We acknowledge Matteo Castagnola and Sander Roet for insightful discussions. R.R.R, T.S.H and H.K. acknowledge funding from the Research Council of Norway through FRINATEK Project Nos. 263110 and 275506 and the Marie Sk{\l}odowska-Curie European Training Network “COSINE— COmputational Spectroscopy In Natural sciences and Engineering,” Grant Agreement No. 765739. We acknowledge computing resources through UNINETT Sigma2—the National Infrastructure for High Performance Computing and Data Storage in Norway, through Project No. NN2962k.

\subsection*{Author Contributions}
R.R.R., T.S.H. and H.K. developed the theory. H.K. conceived the project. R.R.R implemented the computational methodology in the eT program. R.R.R and E.R. proposed the investigated applications. R.R.R, T.S.H and E.R. obtained the data. R.R.R prepared a first draft of the paper. E.R. and R.R.R. worked at the graphical part of the manuscript. H.K. and E.R. oversaw the project. All the authors discussed the results and edited the manuscript.

\subsection*{Competing interests}
The authors declare no competing interests.

\bibliographystyle{mystyle}
\bibliography{references}

\clearpage


\section*{Supplementary material}

# Supplementary Informations

Rosario Roberto Riso[1], Tor S. Haugland[1], Enrico Ronca[2], and Henrik Koch[1,3]

[1]Department of Chemistry, Norwegian University of Science and Technology, 7491 Trondheim, Norway
[2]Istituto per i Processi Chimico Fisici del CNR (IPCF-CNR), Via G. Moruzzi, 1, 56124, Pisa, Italy
[3]Scuola Normale Superiore, Piazza dei Cavalieri 7, 56126 Pisa, Italy

## 1  SC-QED-HF theory

The light-matter interaction is modeled using the Pauli-Fierz dipole Hamiltonian in length gauge

$$H = H_e + \omega b^\dagger b + \lambda\sqrt{\frac{\omega}{2}}(\mathbf{d}\cdot\boldsymbol{\epsilon})(b+b^\dagger) + \frac{\lambda^2}{2}(\mathbf{d}\cdot\boldsymbol{\epsilon}), \tag{1}$$

where the electronic Hamiltonian is given by the expression

$$H_e = \sum_{pq} h_{pq} E_{pq} + \frac{1}{2}\sum_{pqrs} g_{pqrs} e_{pqrs}. \tag{2}$$

In the infinite coupling limit $\lambda \longrightarrow \infty$, the electron-photon interaction dominates the electronic contributions. We therefore only consider the photonic part of the Hamiltonian when determining the eigenfunctions

$$H_\infty = \omega b^\dagger b - \lambda\sqrt{\frac{\omega}{2}}(\mathbf{d}\cdot\boldsymbol{\epsilon})(b+b^\dagger) + \frac{\lambda^2}{2}(\mathbf{d}\cdot\boldsymbol{\epsilon})^2. \tag{3}$$

Since the dipole operator is the only electronic operator in this Hamiltonian, the electronic part of an eigenfunction will be a Slater determinant in the orbitals basis that diagonalizes the dipole. The form of the eigenfunctions are given as:

$$|\psi\rangle = \exp\left(-\frac{\lambda}{\sqrt{2\omega}}\sum_p (\mathbf{d}\cdot\boldsymbol{\epsilon})_{pp} E_{pp}(b-b^\dagger)\right)\prod_i^{n_{occ}} a_i^\dagger|\text{vac}\rangle\otimes|0\rangle, \tag{4}$$

where $|\text{vac}\rangle$ is the vacuum for the electrons and $|0\rangle$ is the photonic vacuum. Transforming the full Hamiltonian to the dipole basis, we obtain

$$\begin{aligned}\tilde{H} =& \tilde{H}_e + \omega b^\dagger b - \lambda\sqrt{\frac{\omega}{2}}\sum_p (\mathbf{d}\cdot\boldsymbol{\epsilon})_{pp} E_{pp}(b+b^\dagger) + \frac{\lambda^2}{2}\sum_{pq}(\mathbf{d}\cdot\boldsymbol{\epsilon})_{pp}(\mathbf{d}\cdot\boldsymbol{\epsilon})_{qq} e_{ppqq}\\ &+ \frac{\lambda^2}{2}\sum_p (\mathbf{d}\cdot\boldsymbol{\epsilon})_{pp}(\mathbf{d}\cdot\boldsymbol{\epsilon})_{pp} E_{pp},\end{aligned} \tag{5}$$

where $\tilde{H}_e$ is equal to

$$\tilde{H}_e = \sum_{pq} \tilde{h}_{pq} E_{pq} + \frac{1}{2}\sum_{pqrs} \tilde{g}_{pqrs} e_{pqrs}. \tag{6}$$

From Eq.4 we define the SC-QED-HF wave function

$$|\psi\rangle = \exp\left(-\frac{\lambda}{\sqrt{2\omega}}\sum_p \omega_p E_{pp}(b-b^\dagger)\right)\exp(\boldsymbol{\kappa})\prod_i^{n_{occ}} a_i^\dagger|\text{vac}\rangle\otimes|0\rangle, \tag{7}$$

where the creation annihilation operators are in the dipole basis. Using the wave function in Eq. 7 we obtain the following energy:

$$\begin{aligned}
E = & \sum_{pq} \tilde{h}_{pq} D_{pq} \exp\left(-\frac{\lambda^2 (w_p - w_q)^2}{4\omega}\right) \\
& + \frac{\lambda^2}{2} \sum_p D_{pp} \left((\mathbf{d} \cdot \boldsymbol{\epsilon})_{pp} - w_p\right) \left((\mathbf{d} \cdot \boldsymbol{\epsilon})_{pp} - w_p\right) \\
& + \frac{\lambda^2}{2} \sum_{pq} \left(D_{pp} D_{qq} - \frac{D_{pq} D_{qp}}{2}\right) \left((\mathbf{d} \cdot \boldsymbol{\epsilon})_{pp} - w_p\right) \left((\mathbf{d} \cdot \boldsymbol{\epsilon})_{qq} - w_q\right) \\
& + \frac{1}{2} \sum_{pqrs} \tilde{g}_{pqrs} \left(D_{pq} D_{rs} - \frac{D_{ps} D_{rq}}{2}\right) \exp\left(-\frac{\lambda^2 (w_p + w_r - w_q - w_s)^2}{4\omega}\right),
\end{aligned} \quad (8)$$

where we note that the contributions from $\tilde{H}_e$ are scaled with Gaussians factors. To derive Eq.8 we used the following identities:

$$\begin{aligned}
U^\dagger E_{pq} U &= E_{pq} \exp\left(\frac{\lambda (w_p - w_q)}{\sqrt{2\omega}} (b - b^\dagger)\right) \\
U^\dagger a_p^\dagger U &= a_p^\dagger \exp\left(\frac{\lambda w_p}{\sqrt{2\omega}} (b - b^\dagger)\right),
\end{aligned} \quad (9)$$

where the unitary operator $U$ is defined as

$$U = \exp\left(-\frac{\lambda}{\sqrt{2\omega}} \sum_n \omega_n E_{nn} (b - b^\dagger)\right). \quad (10)$$

The gradients of the energy with respect to the density matrix elements and the $\omega_p$ parameters are readily obtained

$$\begin{aligned}
\frac{\partial E}{\partial D_{mn}} = & h_{mn} Q_{mn} + \sum_{rs} (2 g_{mnrs} - g_{msrn}) D_{rs} Q_{mnrs} \\
& + \frac{\lambda^2}{2} \delta_{mn} \left((\mathbf{d} \cdot \boldsymbol{\epsilon})_{mm} - w_m\right)^2 \\
& - \lambda^2 D_{nm} \left((\mathbf{d} \cdot \boldsymbol{\epsilon})_{mm} - w_m\right) \left((\mathbf{d} \cdot \boldsymbol{\epsilon})_{nn} - w_n\right) \\
& + 2\lambda^2 \delta_{mn} \left((\mathbf{d} \cdot \boldsymbol{\epsilon})_{mm} - w_m\right) \sum_q D_{qq} \left((\mathbf{d} \cdot \boldsymbol{\epsilon})_{qq} - w_q\right) \\
\frac{\partial E}{\partial w_m} = & \frac{\lambda^2}{\omega} \sum_q h_{mq} D_{mq} (\omega_q - \omega_m) Q_{mq} - \lambda^2 D_{mm} \left((\mathbf{d} \cdot \boldsymbol{\epsilon})_{mm} - \omega_m\right) \\
& + \frac{\lambda^2}{\omega} \sum_{qrs} g_{mqrs} \left(D_{mq} D_{rs} - \frac{D_{ms} D_{rq}}{2}\right) (w_m + w_r - w_q - w_s) Q_{mqrs} \\
& - \lambda^2 \sum_q \left(D_{qq} D_{mm} - \frac{D_{qm}^2}{2}\right) \left((\mathbf{d} \cdot \boldsymbol{\epsilon})_{qq} - w_q\right),
\end{aligned}$$

where $Q_{pq}$ and $Q_{pqrs}$ are defined as:

$$Q_{pq} = \exp\left(-\frac{\lambda^2 (\omega_p - \omega_q)^2}{4\omega}\right) \quad (11)$$

$$Q_{pqrs} = \exp\left(-\frac{\lambda^2 (\omega_p + \omega_r - \omega_s - \omega_q)^2}{4\omega}\right). \quad (12)$$

At a stationary point where the gradients are equal to zero the following relation holds:

$$
\begin{aligned}
\sum_m \frac{\partial E}{\partial \omega_m} &= -\lambda^2 \sum_{mq} \left( D_{qq} D_{mm} - \frac{D_{mq}^2}{2} \right) ((\mathbf{d} \cdot \boldsymbol{\epsilon})_{qq} - \omega_q) - \lambda^2 \sum_m D_{mm}((\mathbf{d} \cdot \boldsymbol{\epsilon})_{mm} - \omega_m) \\
&= \sum_m D_{mm}((\mathbf{d} \cdot \boldsymbol{\epsilon})_{mm} - \omega_m) = 0,
\end{aligned}
\tag{13}
$$

which implies

$$
\sum_m D_{mm}\omega_m = \langle (\mathbf{d} \cdot \boldsymbol{\epsilon}) \rangle. \tag{14}
$$

## 1.1 Total energy and orbital energy origin invariance

In this section we demonstrate the origin invariance of both the total energy and the individual orbital energies for SC-QED-HF. Assume that we have found an optimal set of parameters $\omega_p = \bar{\omega}_p$, $D_{pq} = \bar{D}_{pq}$, such that

$$
\frac{\partial E(D_{pq}, \omega_p)}{\partial D_{pq}}\bigg|_{\bar{D}_{pq}, \bar{\omega}_p} = 0 \tag{15}
$$

$$
\frac{\partial E(D_{pq}, \omega_p)}{\partial \omega_p}\bigg|_{\bar{D}_{pq}, \bar{\omega}_p} = 0. \tag{16}
$$

If the system is displaced in space by a vector $\mathbf{a}$, the dipole matrix elements will change according to:

$$
(\mathbf{d} \cdot \boldsymbol{\epsilon})_{pq} \longrightarrow (\mathbf{d} \cdot \boldsymbol{\epsilon})_{pq} + \frac{Q_{tot}}{N_e}(\mathbf{a} \cdot \boldsymbol{\epsilon})\delta_{pq}, \tag{17}
$$

where $N_e$ is the number of electrons and $Q_{tot}$ is the total charge of the system. Therefore the energy expression in Eq.8 takes the form

$$
\begin{aligned}
E &= \sum_{pq} h_{pq} \bar{D}_{pq} \exp\left( -\frac{\lambda^2 (\bar{w}_p - \bar{w}_q)^2}{4\omega} \right) \\
&+ \frac{\lambda^2}{2} \sum_p \bar{D}_{pp} \left( (\mathbf{d} \cdot \boldsymbol{\epsilon})_{pp} + \frac{Q_{tot}}{N_e}(\mathbf{a} \cdot \boldsymbol{\epsilon}) - \bar{w}_p \right) \left( (\mathbf{d} \cdot \boldsymbol{\epsilon})_{pp} + \frac{Q_{tot}}{N_e}(\mathbf{a} \cdot \boldsymbol{\epsilon}) - \bar{w}_p \right) \\
&+ \frac{1}{2} \sum_{pqrs} g_{pqrs} \left( \bar{D}_{pq} \bar{D}_{rs} - \frac{\bar{D}_{ps} \bar{D}_{rq}}{2} \right) \exp\left( -\frac{\lambda^2 (\bar{w}_p + \bar{w}_r - \bar{w}_q - \bar{w}_s)^2}{4\omega} \right) \\
&+ \frac{\lambda^2}{2} \sum_{pq} \left( \bar{D}_{pp} \bar{D}_{qq} - \frac{\bar{D}_{pq} \bar{D}_{qp}}{2} \right) \left( (\mathbf{d} \cdot \boldsymbol{\epsilon})_{pp} + \frac{Q_{tot}}{N_e}(\mathbf{a} \cdot \boldsymbol{\epsilon}) - \bar{w}_p \right) \left( (\mathbf{d} \cdot \boldsymbol{\epsilon})_{qq} + \frac{Q_{tot}}{N_e}(\mathbf{a} \cdot \boldsymbol{\epsilon}) - \bar{w}_q \right).
\end{aligned}
\tag{18}
$$

After the displacement, the parameters $\bar{\omega}_p$ and $\bar{D}_{pq}$ are not optimal anymore as the energy is changed due to the $\mathbf{a}$ contributions in the dipole. However, we note that the minimum energy is obtained for $\tilde{\omega}_p = \bar{\omega}_p - (Q_{tot}/N_e)(\mathbf{a} \cdot \boldsymbol{\epsilon})$ and the same $\bar{D}_{pq}$ as before. This also implies that the gradients return to zero. Using a similar line of argument we show that the orbital energies are not affected by displacement of the molecule. The dipole integrals that enter the $D_{pq}$ gradient are affected by a displacement of the system as shown in Eq.17. However, the quantity $((\mathbf{d} \cdot \boldsymbol{\epsilon})_{pp} - \omega_p)$ is unchanged choosing $\tilde{\omega}_p$. Therefore, the Fock matrix is unaltered and the eigenvalues remain the same.

## 1.2 Size-extensivity in SC-QED-HF

In this section we show that the SC-QED-HF method is size-extensive.
Assume that we solved the SC-QED-HF equations for two systems A and B

$$
\begin{aligned}
\text{A system:} \quad & |\psi\rangle_A = \exp\left( -\frac{\lambda}{\sqrt{2\omega}} \sum_{p_A} \bar{\omega}_{p_A} E_{p_A p_A}(b - b^\dagger) \right) \exp(\bar{\boldsymbol{\kappa}}_A) \prod_{i_A}^{n_{occ}^A} a_{i_A} |0\rangle_A \\
\text{B system:} \quad & |\psi\rangle_B = \exp\left( -\frac{\lambda}{\sqrt{2\omega}} \sum_{p_B} \bar{\omega}_{p_B} E_{p_B p_B}(b - b^\dagger) \right) \exp(\bar{\boldsymbol{\kappa}}_B) \prod_{i_B}^{n_{occ}^B} a_{i_B} |0\rangle_B
\end{aligned}
\tag{19}
$$

and that their individual energies are $E_A$ and $E_B$. We show here that the product $|\psi_A\psi_B\rangle$ is a stationary point for the AB system. Since $D_{p_A q_B} = 0$ for the product wave function and Eq.14 is respected for A and B separately, the Fock and the $\omega_p$ gradients are equal to

$$\frac{\partial E}{\partial D_{m_A n_A}} = h_{m_A n_A} Q_{m_A n_A} + \sum_{r_A s_A} \left(2 g_{m_A n_A r_A s_A} - g_{m_A s_A r_A n_A}\right) D_{r_A s_A} Q_{m_A n_A r_A s_A}$$

$$+ \frac{\lambda^2}{2} \delta_{m_A n_A} \left((\mathbf{d}\cdot\boldsymbol{\epsilon})_{m_A m_A} - w_{m_A}\right)^2$$

$$- \lambda^2 D_{n_A m_A} \left((\mathbf{d}\cdot\boldsymbol{\epsilon})_{m_A m_A} - w_{m_A}\right)\left((\mathbf{d}\cdot\boldsymbol{\epsilon})_{n_A n_A} - w_{n_A}\right)$$

$$\frac{\partial E}{\partial D_{m_A n_B}} = \frac{\partial E}{\partial D_{n_B m_A}} = 0 \tag{20}$$

$$\frac{\partial E}{\partial D_{m_B n_B}} = h_{m_B n_B} Q_{m_B n_B} + \sum_{r_B s_B} \left(2 g_{m_B n_B r_B s_B} - g_{m_B s_B r_B n_B}\right) D_{r_B s_B} Q_{m_B n_B r_B s_B}$$

$$+ \frac{\lambda^2}{2} \delta_{m_B n_B} \left((\mathbf{d}\cdot\boldsymbol{\epsilon})_{m_B m_B} - w_{m_B}\right)^2$$

$$- \lambda^2 D_{n_B m_B} \left((\mathbf{d}\cdot\boldsymbol{\epsilon})_{m_B m_B} - w_{m_B}\right)\left((\mathbf{d}\cdot\boldsymbol{\epsilon})_{n_B n_B} - w_{n_B}\right) \tag{21}$$

$$\frac{\partial E}{\partial w_{m_A}} = \frac{\lambda^2}{\omega} \sum_{q_A} h_{m_A q_A} D_{m_A q_A} (\omega_{q_A} - \omega_{m_A}) Q_{m_A q_A} - \lambda^2 D_{m_A m_A} \left((\mathbf{d}\cdot\boldsymbol{\epsilon})_{m_A m_A} - \omega_{m_A}\right) \tag{22}$$

$$+ \frac{\lambda^2}{\omega} \sum_{q_A r_A s_A} g_{m_A q_A r_A s_A} \left(D_{m_A q_A} D_{r_A s_A} - \frac{D_{m_A s_A} D_{r_A q_A}}{2}\right) (w_{m_A} + w_{r_A} - w_{q_A} - w_{s_A}) Q_{m_A q_A r_A s_A}$$

$$+ \lambda^2 \sum_{q_A} \frac{D^2_{q_A m_A}}{2} \left((\mathbf{d}\cdot\boldsymbol{\epsilon})_{q_A q_A} - w_{q_A}\right) \tag{23}$$

$$\frac{\partial E}{\partial w_{m_B}} = \frac{\lambda^2}{\omega} \sum_{q_B} h_{m_B q_B} D_{m_B q_B} (\omega_{q_B} - \omega_{m_B}) Q_{m_B q_B} - \lambda^2 D_{m_B m_B} \left((\mathbf{d}\cdot\boldsymbol{\epsilon})_{m_B m_B} - \omega_{m_B}\right) \tag{24}$$

$$+ \frac{\lambda^2}{\omega} \sum_{q_B r_B s_B} g_{m_B q_B r_B s_B} \left(D_{m_B q_B} D_{r_B s_B} - \frac{D_{m_B s_B} D_{r_B q_B}}{2}\right) (w_{m_B} + w_{r_B} - w_{q_B} - w_{s_B}) Q_{m_B q_B r_B s_B}$$

$$+ \lambda^2 \sum_{q_B} \frac{D^2_{q_B m_B}}{2} \left((\mathbf{d}\cdot\boldsymbol{\epsilon})_{q_B q_B} - w_{q_B}\right), \tag{25}$$

which are all equal to zero if Eqs.19 are respected. The energy of the total system is moreover equal to the sum of the energy of the two systems as shown below:

$$
\begin{aligned}
E &= \sum_{p_A q_A} h_{p_A q_A} D_{p_A q_A} Q_{p_A q_A} + \sum_{p_B q_B} h_{p_B q_B} D_{p_B q_B} Q_{p_B q_B} \\
&+ \frac{\lambda^2}{2} \sum_{p_A} D_{p_A p_A} (d_{p_A p_A} - w_{p_A})(d_{p_A p_A} - w_{p_A}) + \frac{\lambda^2}{2} \sum_{p_B} D_{p_B p_B} (d_{p_B p_B} - w_{p_B})(d_{p_B p_B} - w_{p_B}) \\
&+ \frac{\lambda^2}{2} \sum_{p_A q_A} \left( D_{p_A p_A} D_{q_A q_A} - \frac{D_{p_A q_A} D_{q_A p_A}}{2} \right) (d_{p_A p_A} - w_{p_A})(d_{q_A q_A} - w_{q_A}) \\
&+ \frac{\lambda^2}{2} \sum_{p_B q_B} \left( D_{p_B p_B} D_{q_B q_B} - \frac{D_{p_B q_B} D_{q_B p_B}}{2} \right) (d_{p_B p_B} - w_{p_B})(d_{q_B q_B} - w_{q_B}) \\
&+ \frac{1}{2} \sum_{p_A q_A r_A s_A} g_{p_A q_A r_A s_A} \left( D_{p_A q_A} D_{r_A s_A} - \frac{D_{p_A s_A} D_{r_A q_A}}{2} \right) Q_{p_A q_A r_A s_A} \\
&+ \frac{1}{2} \sum_{p_B q_B r_B s_B} g_{p_B q_B r_B s_B} \left( D_{p_B q_B} D_{r_B s_B} - \frac{D_{p_B s_B} D_{r_B q_B}}{2} \right) Q_{p_B q_B r_B s_B} \\
&= E_A + E_B,
\end{aligned}
\tag{26}
$$

where $Q_{pq}$ and $Q_{pqrs}$ are defined as in Eq.12. In passing, we point out that non size-extensive effects in optical cavities can be described using the SC-QED-HF formalism by optimizing the correlation basis or by including more photons in the reference function. These aspects will be discussed and analyzed in a forthcoming publication.

## 2 Correlation in SC-QED-HF

Comparing the QED-HF and the SC-QED-HF wave functions

$$
\begin{aligned}
|\psi\rangle_{\text{QED-HF}} &= \exp\left(-\frac{\lambda \langle \mathbf{d} \cdot \boldsymbol{\epsilon} \rangle}{\sqrt{2\omega}}(b - b^\dagger)\right) \exp\left(\boldsymbol{\kappa}\right) \prod_i^{n_{occ}} a_i^\dagger |vac\rangle \otimes |0\rangle \\
|\psi\rangle_{\text{SC-QED-HF}} &= \exp\left(-\frac{\lambda}{\sqrt{2\omega}} \sum_p \omega_p E_{pp}(b - b^\dagger)\right) \exp\left(\boldsymbol{\kappa}\right) \prod_i^{n_{occ}} a_i^\dagger |vac\rangle \otimes |0\rangle,
\end{aligned}
\tag{27}
$$

we observe that the latter includes explicit electron-photon correlation. In particular, electron-photon correlation is introduced by:

- Allowing every molecular orbital to have a different coherent state $\omega_p$;
- The orbital specific coherent state transformation is in the dipole basis rather that in the canonical basis.

To illustrate how such features influence the method performance, we compare in Fig.1 different QED approaches. In Fig.1a we show how the energy of a water molecule changes inside the cavity due to variations of the coupling . In particular, we compare the results obtained using QED-HF, SC-QED-HF, QED-CCSD-1 and QED-HF-DIP. The latter is obtained from SC-QED-HF enforcing the $\omega_p$ parameters to be equal to the dipole eigenvalues. The energies obtained from QED-HF-DIP increase too fast with the coupling and the optimization of the $\omega_p$ is needed to obtain a good approximation of the wave function. The energy dispersions obtained using QED-HF and SC-QED-HF are more similar to QED-CCSD-1, used here as a reference. Specifically, SC-QED-HF is consistently better than QED-HF.

One of the most important properties of SC-QED-HF is the energy dependence on the cavity frequency. This feature is displayed in Fig.1b for pyrrole. We observe that the SC-QED-HF is in qualitative agreement with the one obtained with QED-CCSD-1, while for QED-HF there is no dispersion at all. Moreover, we note that the agreement becomes increasingly better as the frequency of the cavity is increased because less photons are needed to describe the wave function.

In Fig.1c, the dispersion of the photonic energy $\langle \omega b^\dagger b \rangle$ in length gauge is computed with different methods. As

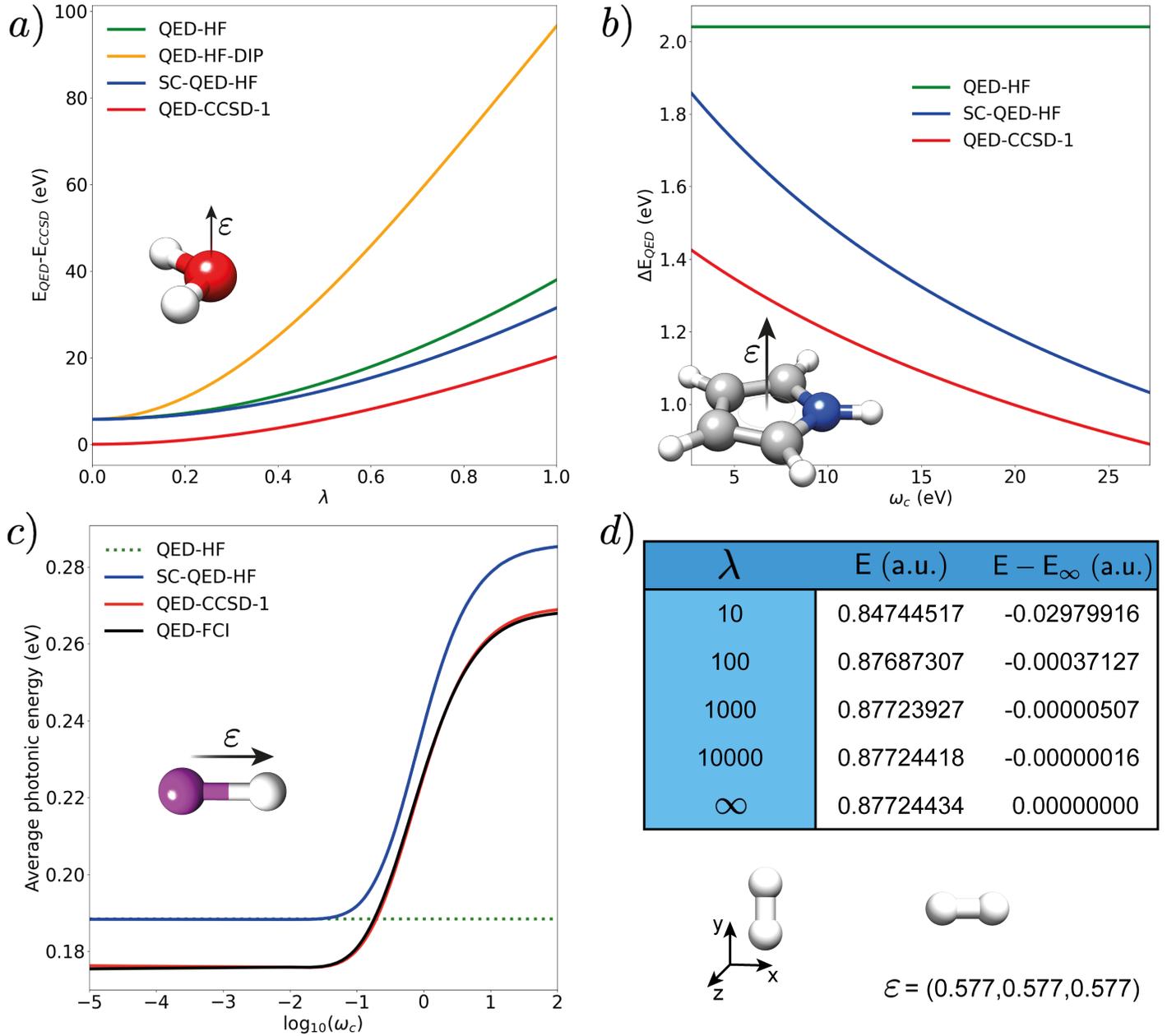

Figure 1: **Properties of the SC-QED-HF method** a) Energy dispersion of a water molecule with respect to the coupling. We notice that SC-QED-HF outperforms all the other HF based methods and shows a coupling dispersion similar to one from QED-CCSD-1. b) Energy dispersion of pyrrole with respect to the cavity frequency. The SC-QED-HF results are qualitatively correct when compared to the QED-CCSD-1 results. c) Dispersion of the photonic energy $\langle \omega b^\dagger b \rangle$. Again, the SC-QED-HF description is qualitatively correct, while a quantitative agreement with the exact solution (QED-FCI) is achieved only using a QED-CCSD-1. d) Convergence of QED-FCI and SC-QED-HF energies for increasing coupling. This table shows that as $\lambda$ increases the QED-FCI energy ( in atomic units ) for two hydrogen molecules converges to the SC-QED-HF energy.

expected, QED-HF does not show any dispersion at all in the frequency while the SC-QED-HF description is qualitatively correct. Quantitative agreement to the exact results (QED-FCI) is only obtained with correlated methods like QED-CCSD-1.

In Fig.1d we show the convergence between the results for very large couplings computed with SC-QED-HF and QED-FCI in this way demonstrating that the SC-QED-HF solution is indeed exact for $\lambda \longrightarrow \infty$.

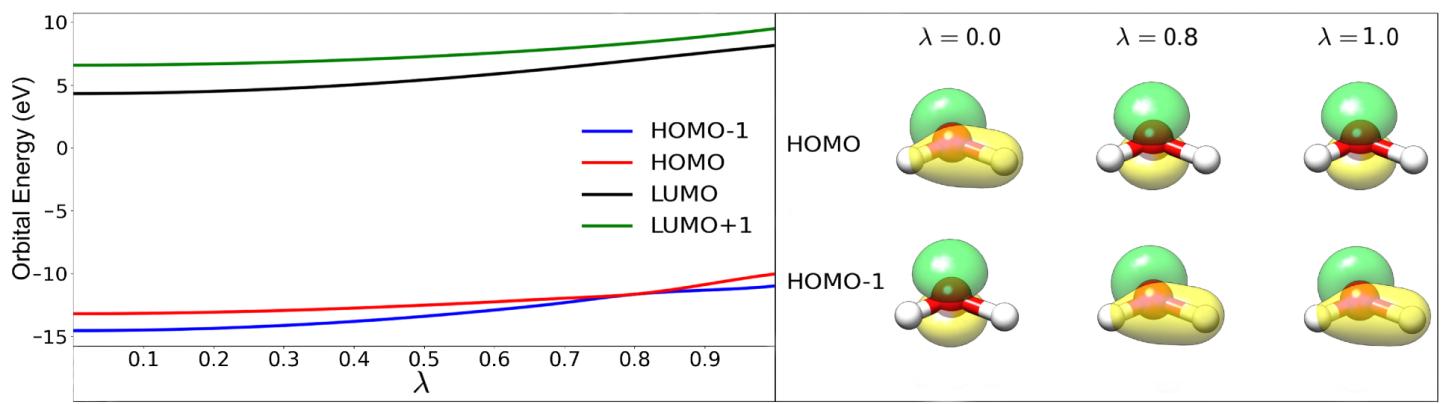

Figure 2: **Symmetry allowed crossing between HOMO and HOMO-1 of a water molecule in an optical cavity**

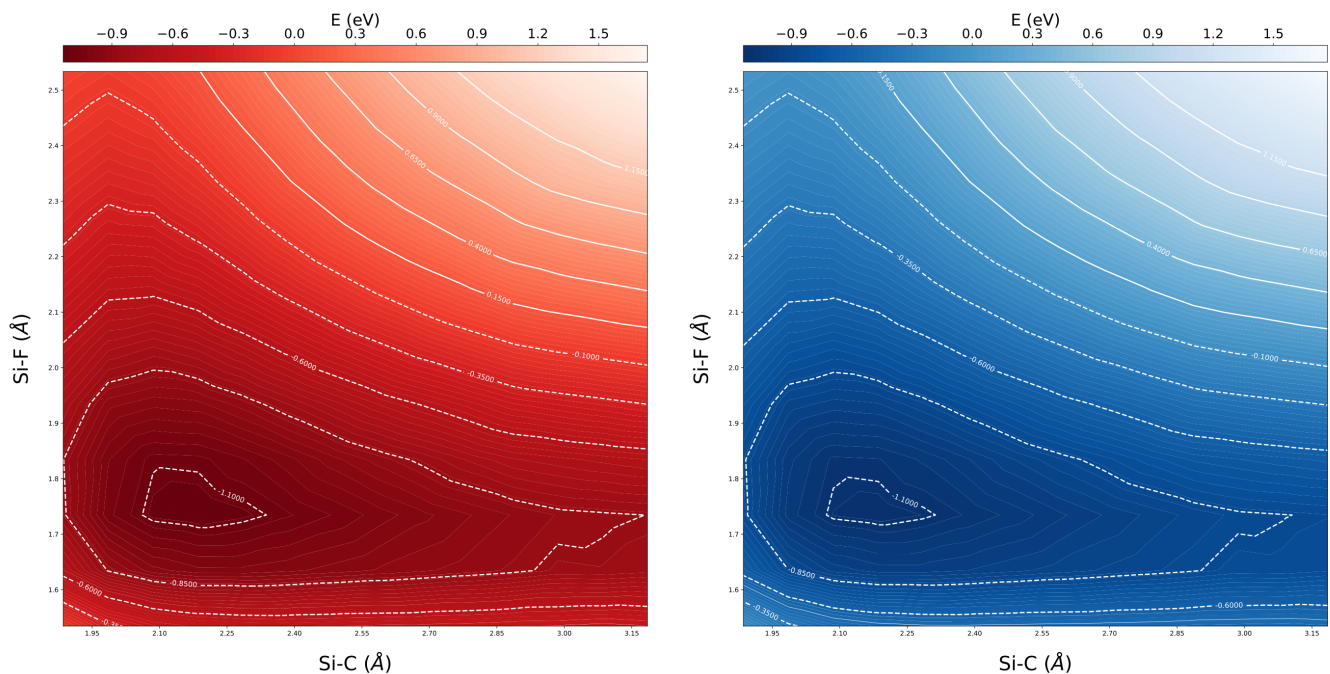

Figure 3: **Potential energy surfaces for the TPA deprotection in TBAF inside and outside the cavity.** The two surfaces have been shifted such that the point (1.9Å, 2.5Å) has zero energy. The PES shape is not affected significantly by the presence of the cavity.

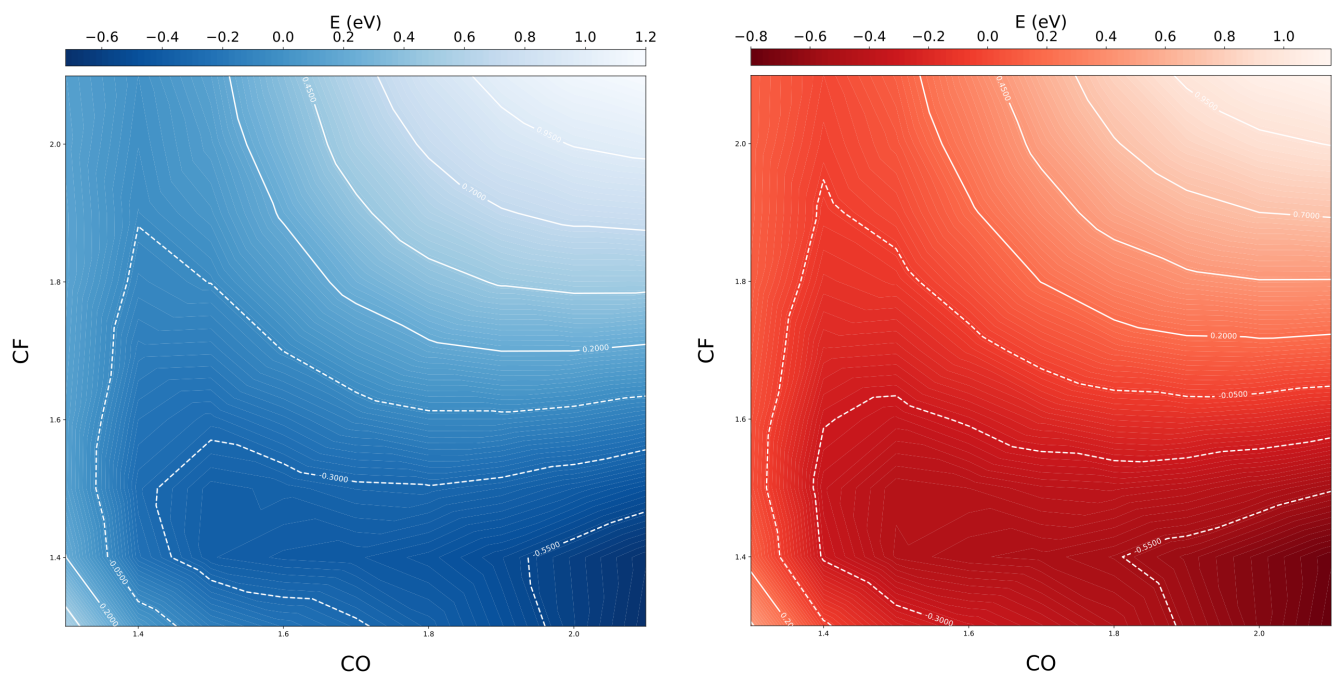

Figure 4: **Potential energy surface for the solvolysis of PNPA inside and outside the cavity.** The two surfaces have been shifted such that the point (1.3Å, 2.1Å) has zero energy. The PES shape is not affected significantly by the presence of the cavity.


\end{document}